\pdfoutput=1

\documentclass[11pt,twoside,a4paper,cmspaper,final,collab]{cms-tdr}
\def\svnVersion{141593a-D}\def\svnDate{2020/11/09}\def\cmsCernNoTag{CERN-EP-2020-075}\def\cmsCernDate{\today}\def\cmsMessage{"Published in the Journal of High Energy Physics as \href{http://dx.doi.org/10.1007/JHEP11(2020)001}{\doi{10.1007/JHEP11(2020)001}.}"}

\begin{document}\cmsNoteHeader{BPH-14-009}

\hyphenation{had-ron-i-za-tion}
\hyphenation{cal-or-i-me-ter}
\hyphenation{de-vices}
\RCS$HeadURL: svn+ssh://svn.cern.ch/reps/tdr2/papers/BPH-14-009/trunk/BPH-14-009.tex $
\RCS$Id: BPH-14-009.tex 477977 2018-10-13 17:18:45Z tzie $

\newcommand{\PgUn}{\PGUP{nS}}
\newcommand{\RapDimuon}{\ensuremath{\abs{y^{\PGm\PGm}}}\xspace}
\newcommand{\PtDimuon}{\ensuremath{\pt^{\PGm\PGm}}\xspace}
\newcommand{\MeanPtDimuon}{\ensuremath{\bigl\langle\pt^{\PGm\PGm}\bigr\rangle}\xspace}
\newcommand{\EtaTrack}{\ensuremath{\abs{\eta^{\text{track}}}}\xspace}
\newcommand{\PtTrack}{\ensuremath{\pt^{\text{track}}}\xspace}
\newcommand{\Multiplicity}{\ensuremath{N_{\text{track}}}\xspace}
\newcommand{\MultiplicityDir}{\ensuremath{N_{\text{track}}^{\Delta\phi}}\xspace}
\newcommand{\MultiplicityCone}{\ensuremath{N_{\text{track}}^{\Delta R}}\xspace}
\newcommand{\MultiplicitynoSpace}{\ensuremath{N_{\text{track}}}}
\newcommand{\MultiplicitynoSpaceMean}{\ensuremath{\bigl\langle N_{\text{track}} \bigr\rangle}}
\newcommand{\MultiplicityTrue}{\ensuremath{N^{\text{true}}_{\text{track}}}\xspace}
\newcommand{\MultiplicityMean}{\ensuremath{\bigl\langle N_{\text{track}} \bigr\rangle }\xspace}
\newcommand{\MultiplicityMeanDir}{\ensuremath{\bigl\langle N_{\text{track}}^{\Delta\phi} \bigr\rangle }\xspace}
\newcommand{\PtMu}{\ensuremath{\pt^{\PGm}}\xspace}
\newcommand{\EtaMu}{\ensuremath{\abs{\eta^{\PGm}}}\xspace}
\newcommand{\ST}{\ensuremath{S_{\mathrm{T}}}\xspace}
\newcommand{\RatioA}{\ensuremath{\PgUb/\PgUa}\xspace}
\newcommand{\RatioB}{\ensuremath{\PgUc/\PgUa}\xspace}

\providecommand{\cmsTable}[1]{\resizebox{\textwidth}{!}{#1}}

\cmsNoteHeader{BPH-14-009} 

\title{Investigation into the event-activity dependence of \PGUP{nS} relative production in proton-proton collisions at $\sqrt{s} = 7\TeV$}

\date{\today}

\abstract{
The ratios of the production cross sections between the excited \PgUb and \PgUc mesons and the \PgUa ground state, detected via their decay into two muons, are studied as a function of the number of charged particles in the event. The data are from proton-proton collisions at $\sqrt{s} = 7\TeV$, corresponding to an integrated luminosity of 4.8\fbinv, collected with the CMS detector at the LHC. Evidence of a decrease in these ratios as a function of the particle multiplicity is observed, more pronounced at low transverse momentum \PtDimuon. For \PgUn mesons with $\PtDimuon > 7\GeV$, where most of the data were collected, the correlation with multiplicity is studied as a function of the underlying event transverse sphericity and the number of particles in a cone around the \PgUn direction. The ratios are found to be multiplicity independent for jet-like events. The mean \PtDimuon values for the \PgUn states as a function of particle multiplicity are also measured and found to grow more steeply as their mass increases.
}

\hypersetup{
pdfauthor={CMS Collaboration},
pdftitle={Investigation into the event-activity dependence of Y(nS) relative production in proton-proton collisions at sqrt(s) = 7 TeV},
pdfsubject={CMS},
pdfkeywords={CMS, physics, b quark}}

\maketitle

\section{Introduction}
A wealth of experimental data on quarkonium production is available~\cite{Andronic:2015wma}, but very little of it investigates the relationship to the underlying event (UE).  For instance, the fragmentation of soft gluons~\cite{Kramer:2001hh} or feed-down processes~\cite{Digal:2001ue}~(decays of higher-mass states to a lower-mass one), could generate different numbers of particles associated with each of the quarkonium states. Therefore, the global event characteristics (multiplicity, sphericity, etc.) may show variations that depend on the quarkonium state.
Recent observations in proton-proton ($\Pp\Pp$) collisions at the LHC have shown that \PJGy~\cite{Abelev:2012rz} and \PD~\cite{Adam:2016ich} meson yields increase with the associated track multiplicity, which has been explained as a consequence of multiparton interactions~\cite{Thakur:2017kpv}. The same effect was seen in $\Pp\Pp$ and proton-lead ($\Pp \mathrm{Pb}$) collisions~\cite{Chatrchyan:2013nza} for \PgUn mesons, where $\mathrm{n} = (1, 2, 3)$, with the additional observation that this effect is more pronounced for the ground state than for the excited states. 

A host of  results obtained in $\Pp\Pp$ collisions at the LHC~\cite{Khachatryan:2010gv,Aad:2015gqa,Khachatryan:2015lva,Adam:2016emw,Khachatryan:2016txc,Khachatryan:2016yru}
may be interpreted as a signal of collective effects in the high particle density environment created at TeV energies~\cite{Campanini:2011bj,Dusling:2015gta}. However, it is still not clear whether the small-size system created in $\Pp\Pp$ collisions could exhibit fluid-like properties due to early  thermalisation, as observed in PbPb collisions~\cite{NatureTerm,Nagle:2018nvi}. Some of the collective effects detected so far could possibly be reproduced by fragmentation of saturated gluon states~\cite{Schenke:2016lrs} or by the Lund string model~\cite{Ferreres-Sole:2018vgo}. These observations suggest that different phenomena need to be considered for a full understanding of the quarkonium and heavy-flavour production mechanisms.  An analysis of the dependence of quarkonium yields as a function of the number of charged particles produced in the event in $\Pp\Pp$  collisions may help to resolve some of these questions~\cite{Lang:2013ex,Ferreiro:2012fb}, in particular in interpreting the observed production rates in heavy ion collisions~\cite{Sirunyan:2018nsz}. 
 
In this paper, measurements are presented of the cross section ratios, multiplied by the branching fractions to a muon pair~\cite{PDG2018}, of the bottomonium excited states \PgUb and \PgUc to the ground state \PgUa (indicated by \RatioA and \RatioB, respectively)  as a function of the number of charged particles per event in $\Pp\Pp$ collisions at a centre-of-mass energy of $\sqrt{s} = 7\TeV$. 

The data were collected in 2011 by the CMS experiment at the LHC. The \PgUn states are detected via their dimuon decay in the \PgUn  rapidity range  $\RapDimuon < 1.2$. The charged particle multiplicity of the interaction containing the dimuon, \Multiplicity, is calculated starting from the number of reconstructed tracks with transverse momentum $\PtTrack > 0.4\GeV$ and pseudorapidity $\EtaTrack < 2.4$, and correcting  for the track reconstruction efficiency.
Together with the \PgUn cross section ratios, the evolution of the average transverse momentum of the \PgU~states, \MeanPtDimuon, is studied with respect to \Multiplicity. 
For $\PtDimuon> 7\GeV$, additional observables are considered to characterise the dependence of the production cross section ratios on \Multiplicity, including the  number of particles produced in various angular regions with respect to the \PgUn momentum direction, the number of particles in a restricted cone around this direction, and the transverse sphericity of charged particles in the event.

\section{The CMS detector}
\label{sec:cms}

The central feature of the CMS apparatus is a superconducting solenoid of 6\unit{m} internal diameter, providing a magnetic field of 3.8\unit{T}. Within the solenoid volume are a silicon pixel and strip tracker, a lead tungstate crystal electromagnetic calorimeter, and a brass and scintillator hadron calorimeter, each composed of a barrel and two endcaps sections. Forward calorimeters extend the $\eta$ coverage provided by the barrel and endcap detectors. Muons are detected in gas-ionisation chambers embedded in the steel flux-return yoke outside the solenoid. 

The silicon tracker measures charged particles within the range $\EtaTrack < 2.5$. During the LHC running period when the data used in this paper were recorded, the silicon tracker consisted of 1440 silicon pixel and 15\,148 silicon strip detector modules. For nonisolated particles of $1 < \PtTrack < 10\GeV$ and $\EtaTrack < 1.4$, the track resolutions are typically 1.5\% in \PtTrack and 25--90 (45--150)\mum in the transverse (longitudinal) impact parameter~\cite{TRK-11-001}.

Muons are measured in the range $\EtaMu < 2.4$, with detection planes made using three technologies: drift tubes, cathode strip chambers, and resistive plate chambers. Matching muons to tracks measured in the silicon tracker results in a transverse momentum resolution between 1\% and 2.8\%, for \PtMu up to 100 \GeV~\cite{Chatrchyan:2012xi}.  

Events of interest are selected using a two-tiered trigger system~\cite{Khachatryan:2016bia}. The first level, composed of custom hardware processors, uses information from the calorimeters and muon detectors to select events at a rate of around 100\unit{kHz} within a time interval of less than 4\mus. The second level, known as the high-level trigger, consists of a farm of processors running a version of the full event reconstruction software optimised for fast processing, and reduces the event rate to around 1\unit{kHz} before data storage. 

A more detailed description of the CMS detector, together with a definition of the coordinate system used and the relevant kinematic variables, can be found in Ref.~\cite{Chatrchyan:2008aa}.

\section{Data analysis}
\label{procedure}
\subsection{Event selection}

The trigger used to select events for this analysis requires an opposite-sign muon pair with an invariant mass $8.5 <\mathrm{m}_{\mu\mu}<11.5\GeV$, and $\RapDimuon<1.25$, with no explicit \PT requirement on the muons. Additionally, the dimuon vertex fit $\chi^2$ probability has to be greater than 0.5\% and the distance of closest approach between the two muons less than 5\unit{mm} . Events where the two muons bend toward each other in the magnetic field, such that their trajectory can cross within the muon detectors, are rejected to limit the trigger rate, while retaining the highest quality muon pairs. During the 2011 data taking, the increase in the LHC instantaneous luminosity necessitated the increase of the minimum \PtDimuon requirement  to maintain a constant rate for \PgUn events. The collected data correspond to an integrated luminosity of 0.3\fbinv, 1.9\fbinv, and 4.8\fbinv for minimum \PtDimuon requirements of 0, 5, and 7\GeV, respectively. For the inclusive $\PtDimuon > 0$ sample, the data are weighted according to the relative integrated luminosity of the period in which they were taken.

In the offline analysis, two reconstructed opposite-sign muon tracks~\cite{Chatrchyan:2013sba} are required to match the triggered muons. Each muon candidate must pass a pseudorapidity-dependent \pt requirement with $\PtMu > 2\GeV$ for $1.6 < \EtaMu < 2.4$, $\PtMu > 3.5\GeV$ for $\EtaMu<1.2$, and a linear interpolation of the \PtMu threshold for $1.2 <\EtaMu <1.6$. Given the \RapDimuon trigger constraints, the analysis is restricted to the kinematic region \RapDimuon$< 1.2$. In addition, the muon tracks are each required to have at least 11 tracker hits, including at least two hits in the pixel detector. The track fit must have a $\chi^2$ per degree of freedom (ndf) below 1.8 and the tracks must intersect the beam line within a cylinder of radius 3\unit{cm} and length $\pm$30\unit{cm} around the detector centre. Finally, the $\chi^2$ probability of the vertex fit must exceed 1\%. These selection criteria result in 3 million candidates within the invariant mass range $8.6 < \mathrm{m}_{\PGm\PGm} <11.3\GeV$ used to extract the signal.

\subsection{Track multiplicity evaluation}
In 2011, the average number of reconstructed $\Pp\Pp$ collision vertices per bunch crossing (pileup) was seven. The reconstructed $\Pp\Pp$ collision vertex that is closest to the dimuon vertex is considered as the production vertex (PV), and events in which another vertex is located closer than 0.2\unit{cm} along the beam line are discarded. This removes 8\% of the events. The PV must be located within 10\unit{cm} of the centre of the detector along the beamline, where the track reconstruction efficiency is constant.

The contribution of every  track to the PV is given as a weight~\cite{TRK-11-001}. A track is considered associated if this weight is above 0.5, and the multiplicity is measured by considering the associated tracks that satisfy the high-purity criteria of Ref.~\cite{TRK-11-001}. These criteria use the number of silicon tracker layers with hits, the $\chi^2$/ndf of the track fit, and the impact parameter with respect to the beamline to reduce the number of spurious tracks.
In addition, the following criteria are designed to check the quality of the tracks and  ensure that they emanate from the PV. The transverse and longitudinal impact parameters of each track with respect to the PV must be less than three times the calculated uncertainty in the impact parameter. The tracks must also have a calculated relative \pt uncertainty less than 10\%, $\EtaTrack<2.4$, and $\PtTrack>0.4\GeV$. The muon tracks are used in the vertex reconstruction, but are not counted in \Multiplicity.

Detector effects in track reconstruction are studied with Monte Carlo (MC) samples generated  with \PYTHIA 8.205~\cite{Sjostrand:2014zea} and a UE tune CUETP8M1~\cite{Khachatryan:2015pea},  using a full simulation of the CMS detector response based on \GEANTfour~\cite{Agostinelli:2002hh}.
The MC samples are reconstructed with the same software framework used for the data, including an emulation of the trigger. The track reconstruction efficiency for tracks originating from the PV and within the chosen kinematic region increases from 60\% at $\PtTrack = 0.4\GeV$ to greater than 90\% for $\PtTrack > 1\GeV$, with an average value of 75\%. The rate of misreconstructed tracks  (tracks coming from the reconstruction algorithms not matched with a simulated track) is 1--2\%. Following the method of Ref.~\cite{Chatrchyan:2013ala}, two-dimensional maps in \EtaTrack and \PtTrack of the tracker efficiency and misreconstruction rate, are used to produce a factor for each track, given by the complement to 1 of the misreconstruction rate, divided by the efficiency. The \Multiplicity value is given by the sum of the associated tracks weighted by this factor. To evaluate the systematic uncertainties in the track multiplicity, correction maps are produced using different types of processes (such as Drell--Yan and multijet events) and another \PYTHIA UE tune (4C~\cite{Corke:2010yf}). The effect on the final \Multiplicity is of the order of 1\%. This is combined in quadrature  with the uncertainty in the tracking efficiency, which is 3.9\% for a single track~\cite{TRK-11-001}. 
In the selected data sample, the mean track \PT is around 1.4\GeV and the mean corrected multiplicity $\MultiplicitynoSpaceMean = 37.7 \pm 0.1\stat \pm 1.4\syst$. This multiplicity is about twice the value of 17.8 found in an analysis of minimum bias (MB) events~\cite{Khachatryan:2010gv}, which do not have any selection bias. The average corrected  multiplicity is  shown for 20 \Multiplicity ranges in Table~\ref{tab:First}. The same binning is used for the \PgUn ratios for $\PtDimuon > 7\GeV$ as a function of \Multiplicity. Different \Multiplicity binning has been used for the other results, to take into account the available event statistics with alternative selections.

While the described \Multiplicity variable is used for all the results in this paper, to facilitate comparisons with theoretical models,  the  corresponding  true  track  multiplicity (\MultiplicityTrue) was also evaluated, where simulated stable charged particles ($c\tau >10\unit{mm}$) are counted. 
A large Drell--Yan \PYTHIA sample was used, which was produced with the same pileup conditions as data.  Given the difference in the \Multiplicity distribution between data and simulation, the simulation events have been reweighed to reproduce the \Multiplicity distribution in data. Then, for every range of \Multiplicity, the  \MultiplicityTrue distribution is  produced both for $\PtTrack >0.4\GeV$ and $> 0\GeV$. These distributions are fitted with two half-Gaussians, which are folded normal distributions having the same mean and different standard deviations on the left and right sides. The most probable values from the fits are listed in  the third and fourth columns of Table~\ref{tab:First} for $\PtTrack > 0.4\GeV$ and 0\GeV, respectively. For $\PtTrack > 0.4\GeV$ the values are similar to those for \MultiplicityMean, except at high multiplicity.  This is due to the probability of merging two nearby vertices during reconstruction, which moves events from low to high multiplicity. Using the same PYTHIA simulation, where a merged vertex can be easily tagged by comparison with the generator-level information, we find that for the 2011 pileup conditions the percentage of merged vertices is below 1\% for \Multiplicity$< 30$, and reaches 13\% in the highest-multiplicity bin. Table~\ref{tab:First} also reports the percentage of background MB events in data for each multiplicity bin.

\begin{table}[h]
\centering
\topcaption{Efficiency-corrected multiplicity bins used in the \PgUn ratio analysis and  the corresponding mean number of charged particle tracks with $\PtTrack> 0.4\GeV$ in the data sample. The most probable values of the two half-Gaussian fit to the corresponding \MultiplicityTrue in simulation, for $\PtTrack>0.4\GeV$ and $\PtTrack> 0\GeV$, are also indicated. The uncertainties shown are statistical, except for  \MultiplicityMean, where the systematic uncertainties are also reported. In the last column, the  percentage of minimum bias (MB) events in the different multiplicity bins is also indicated.}
\cmsTable{
\begin{tabular}{ccccc}
\Multiplicity & \MultiplicityMean  & \MultiplicityTrue $\bigl( \PtTrack > 0.4\GeV \bigr)$  &  \MultiplicityTrue $\bigl( \PtTrack > 0\GeV \bigr)$ & MB (\%)\\
\hline
0--6  	&  4.2		$\pm$   0.2 $\pm$   0.1&   4.2 $\pm$   0.3  &		 6.6 $\pm$    0.6 & 26.94 $\pm$ 0.03\\
6--11	&  8.8 		$\pm$   0.4 $\pm$   0.3&   8.9 $\pm$   0.4  &		 14.9 $\pm$   0.9 & 16.73 $\pm$ 0.03\\
11--15	&  13.1  	$\pm$   0.5 $\pm$   0.4&  13.4 $\pm$   0.4  &		 22.7 $\pm$   0.9 & 10.21 $\pm$ 0.02\\
15--19  	&  17.1	 	$\pm$   0.7 $\pm$   0.6&  17.1 $\pm$   0.4  &		 28.5 $\pm$   0.9 & 8.39 $\pm$ 0.02\\
19--22  	&  20.5	 	$\pm$   0.8 $\pm$   0.7&  20.7 $\pm$   0.4  &		 35.4 $\pm$   1.0 & 5.36 $\pm$ 0.02\\
22--25  	&  23.5	 	$\pm$   0.9 $\pm$   0.8&  23.5 $\pm$   0.4  &		 40.3 $\pm$   1.0 & 4.70 $\pm$ 0.02\\
25--28  	&  26.5	 	$\pm$   1.0 $\pm$   0.9&  26.4 $\pm$   0.4  &		 43.6 $\pm$   1.0 & 4.12 $\pm$ 0.01\\
28--31  	&  29.5	 	$\pm$   1.2 $\pm$   1.0&  29.3 $\pm$   0.5  &	 	 48.5 $\pm$   1.0 & 3.61 $\pm$ 0.01\\
31--34  	&  32.5	 	$\pm$   1.3 $\pm$   1.1&  32.2 $\pm$   0.5  &		 53.0 $\pm$   1.0 & 3.12 $\pm$ 0.01\\
34--37  	&  35.5	 	$\pm$   1.4 $\pm$   1.2&  35.1 $\pm$   0.5  &		 57.6 $\pm$   1.0 & 2.72 $\pm$ 0.01\\
37--40  	&  38.5	 	$\pm$   1.5 $\pm$   1.3&  38.0 $\pm$   0.5  &		 62.1 $\pm$   1.1 & 2.60 $\pm$ 0.01 \\
40--44  	&  42.0	 	$\pm$   1.6 $\pm$   1.4&  41.3 $\pm$   0.5  &		 67.2 $\pm$   1.1 & 2.36 $\pm$ 0.01\\
44--48  	&  45.9	 	$\pm$   1.8 $\pm$   1.5&  45.1 $\pm$   0.6  &		 72.8 $\pm$   1.2 & 2.21 $\pm$ 0.01\\
48--53  	&  50.4	 	$\pm$   2.0 $\pm$   1.7&  49.4 $\pm$   0.6  &		 79.1 $\pm$   1.2 & 2.01 $\pm$ 0.01\\
53--59  	&  55.8	 	$\pm$   2.2 $\pm$   1.9&  54.4 $\pm$   0.6  &		 86.6 $\pm$   1.2 & 1.75 $\pm$ 0.01 \\
59--67  	&  62.7	 	$\pm$   2.5 $\pm$   2.1&  60.8 $\pm$   0.6  &		 95.8 $\pm$   1.3 & 1.41 $\pm$ 0.01\\
67--80  	&  72.6	 	$\pm$   2.9 $\pm$   2.4&  69.6 $\pm$   0.6  &		 109.2 $\pm$   1.3 & 1.12 $\pm$ 0.01\\
80--95  	&  86.0	 	$\pm$   3.4 $\pm$   2.9&  81.9 $\pm$   0.6  &		 126.4 $\pm$   1.4 & 0.459 $\pm$ 0.005\\
95--110 & 100.1		$\pm$   4.0 $\pm$   3.3&  95.8 $\pm$   0.9  &		 145.0 $\pm$   1.6 & 0.121 $\pm$ 0.002\\
110--140 &118.7		$\pm$   4.9 $\pm$   3.9& 109.4 $\pm$   1.2  &		 164.5 $\pm$   2.0 & 0.0038 $\pm$ 0.0001\\
\end{tabular}
}
\label{tab:First}
\end{table}

\subsection{Signal extraction}
In each multiplicity bin listed in Table~\ref{tab:First}, an extended binned maximum likelihood fit is performed on the dimuon invariant mass distribution, using the  \textsc{RooFit} toolkit~\cite{Verkerke:2003ir}. 
Each signal peak is described by functions with a Gaussian core and an exponential tail on the low side. The Gaussian core takes into account the reconstructed dimuon mass resolution, which is much larger than the natural widths of the \PgUn states. The exponential tail describes the effect from final-state radiation.
This function, usually referred to as \textit{GaussExp}~\cite{Das:2016stf}, is continuous in its value  and first derivative. It has two  parameters for the mean and width of the Gaussian function and one parameter for the decay constant of the exponential tail. Each peak is fitted with two \textit{GaussExp} functions, which differ only in the widths of the Gaussians, to describe the \PT and rapidity dependence  of the resolution. The means of the Gaussian functions are constrained to the world-average \PgUn  masses~\cite{PDG2018}, multiplied by a common free factor to take into account the slightly  shifted experimental dimuon mass scale~\cite{Chatrchyan:2012xi}. The  widths of the two Gaussian functions are constrained to scale between the three signal peaks, following the ratios of their world-average  masses. The tail parameter of the exponential is left free in the fit, but is common to the three \PgUn signal shapes. 
There are eight resulting free parameters in the fit: the mass scale factor, the two widths of the \PgUa Gaussian function, their respective fraction in describing the \PgUa peak,  the  tail parameter of the exponential, the number of \PgUa events, and the  ratios \RatioA and \RatioB. The validity of the fit choices, in particular of the fixed mass resolution scaling between the three states, has been confirmed by relaxing these constraints and comparing the results in larger \Multiplicity bins, to decrease the sensitivity to statistical fluctuations. To describe the background, an \emph{Error Function} combined with  an exponential is chosen. 

\begin{figure}[h!]
\centering
   \includegraphics[width=0.45\textwidth]{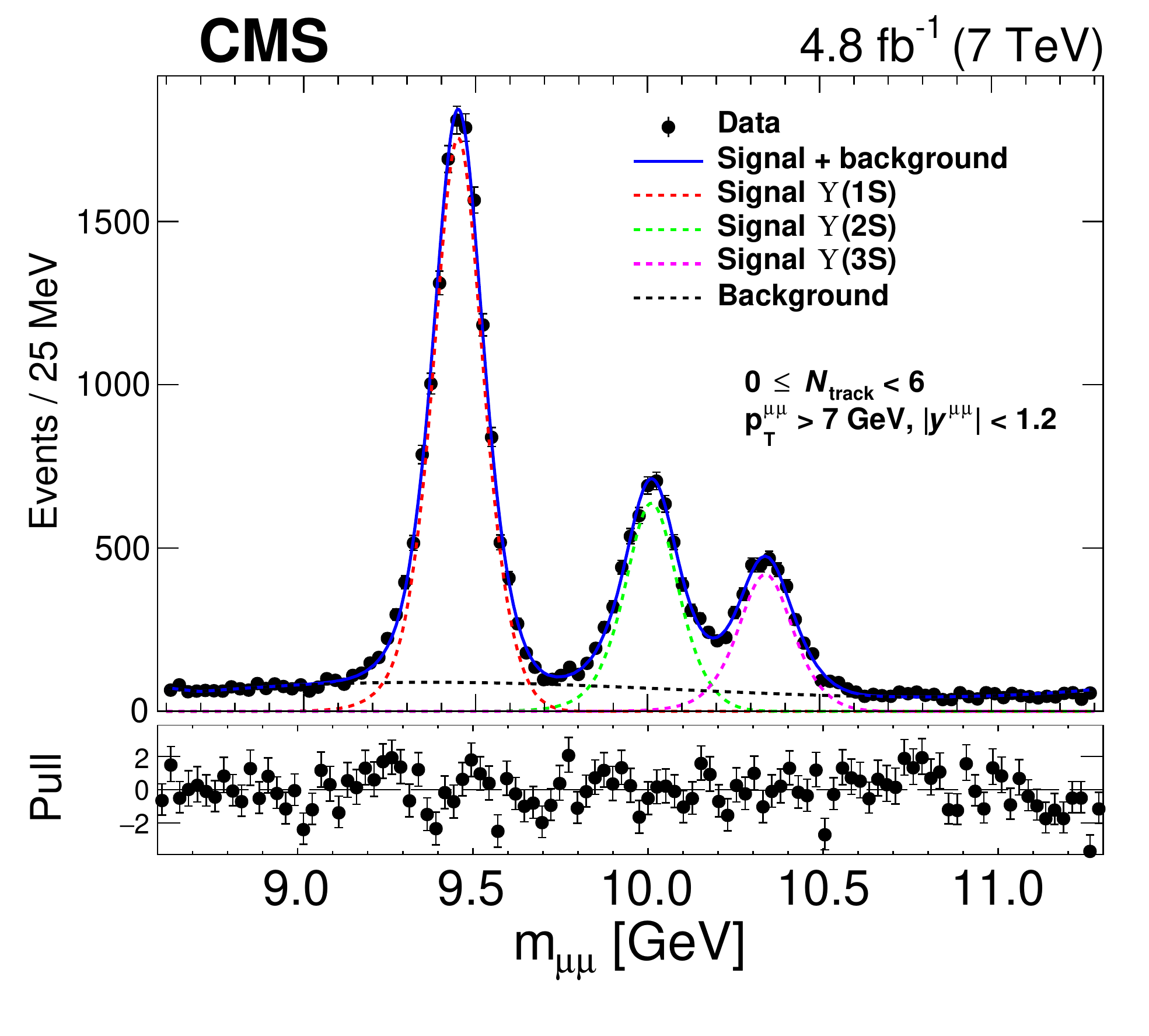}
   \hspace{5mm}
   \includegraphics[width=0.45\textwidth]{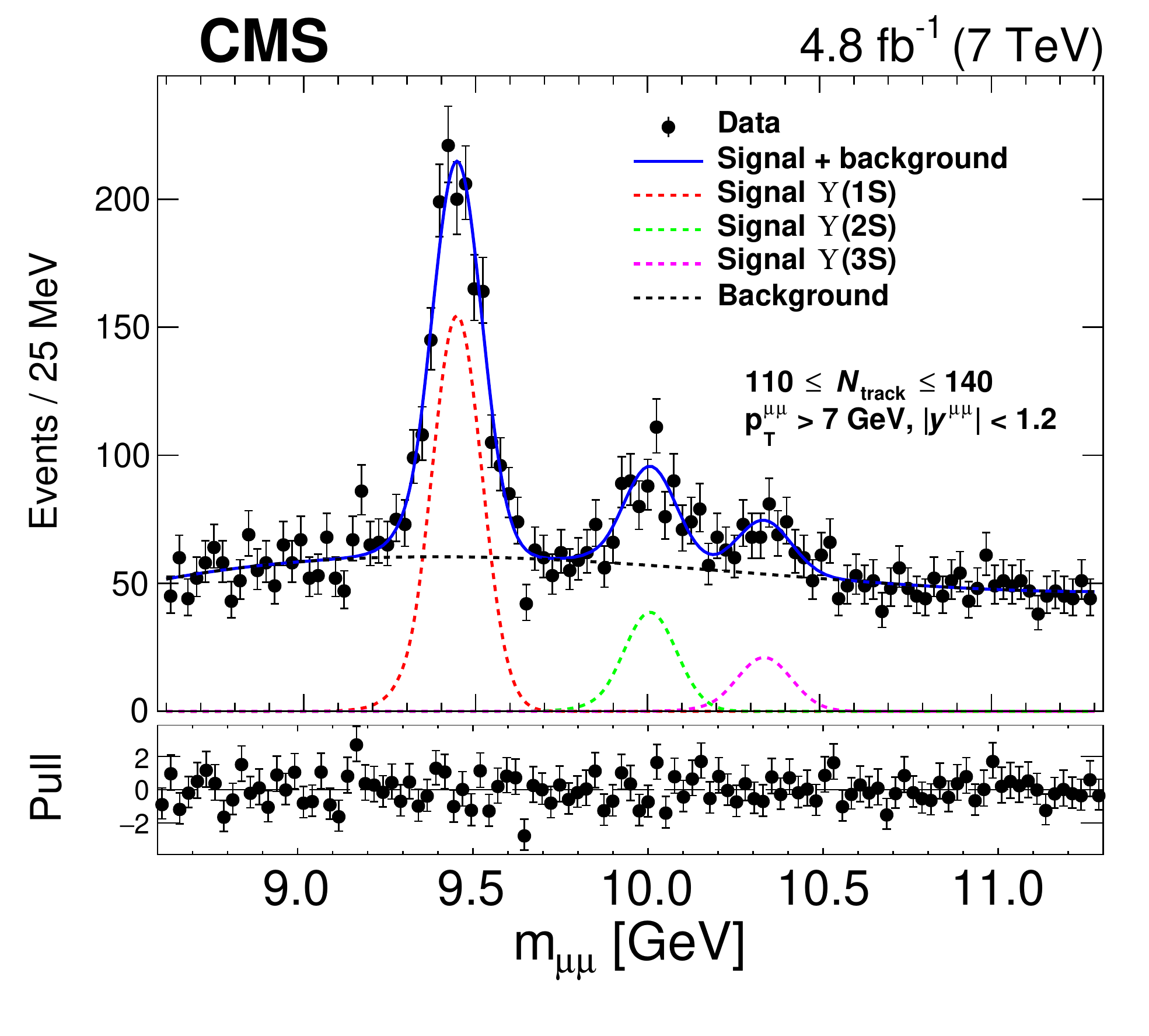}
   \caption{The \PGmp\PGmm invariant mass distributions for dimuon candidates with $\PtDimuon> 7\GeV$ and $\RapDimuon< 1.2$, in two intervals of charged particle multiplicity, 0--6 (left) and 110--140 (right). The result of the fit is shown by the solid lines, with the various dotted lines giving the different components. The lower panel displays the pull distribution.}
\label{fig:fits}
\end{figure}

Examples of the invariant mass distributions and the results of the fit are shown in Fig.~\ref{fig:fits} for $\Multiplicity =$ 0--6 (left) and 110--140 (right). The lower panel displays the normalised residual (pull) distribution. This is given by the difference between the observed number of events in the data and the integral of the fitted signal and background function in that bin, divided by the Poisson statistical uncertainty in the data. The lineshape description represents the data well and  shows no systematic structure.
Signal extraction was found to be the main source of systematic uncertainties in the measurement of the ratios. In order to evaluate it, eight alternative fit functions have been considered, combining the described ones and alternative modelling of the signal (Crystal Ball functions~\cite{CrystalBallRef}) and the background (polynomials of different orders, exponential function). The maximum variation with respect to the chosen fit is taken as the systematic uncertainty, and is found to be up to 5.5\% in the highest \Multiplicity bins.

\subsection{Acceptances, efficiencies and vertex merging corrections}
\label{sec:corr}
Evaluation of the efficiencies begins with the single-muon reconstruction efficiencies obtained with a ''tag-and-probe'' approach~\cite{Khachatryan:2010xn}, based on \PJGy control samples in data. The dimuon efficiency is then obtained by combining the single-muon efficiencies and a factor that takes into account the trigger inefficiency for close-by muons, obtained from  MC simulation, following the procedure detailed in Ref.~\cite{Khachatryan:2015qpa}.

The acceptances for the three upsilon states are evaluated using an unpolarised hypothesis in the \PYTHIA + \EVTGEN 1.4.0p1~\cite{Lange:2001uf} and $\PHOTOS$~3.56~\cite{Barberio:1993qi} packages. This hypothesis was chosen since there is no evidence for large \PgUn polarisation at LHC energies~\cite{Chatrchyan:2012woa}, nor any dependence of the polarisation on multiplicity~\cite{Khachatryan:2016vxr}. No systematic uncertainties are assigned for this assumption. 

While the efficiency is determined event-by-event, the \PtDimuon-dependent acceptance correction is different for the three upsilon states and the background. As a first step a \PtDimuon-dependent distribution for the efficiency is obtained from all the candidates in a considered multiplicity range, associating the calculated \PgUn candidate efficiency to its measured \PtDimuon. Then, the true  \PtDimuon distribution from data is extracted using  the \emph{sPlot}~\cite{Pivk:2004tyr} technique. This method provides an event-by-event weight, based on the value of $\mathrm{m}_{\PGm\PGm}$, that allows us to reconstruct the \PtDimuon distribution, corrected for the background contribution. This experimental \PtDimuon distribution for the three \PgUn~states is rescaled by the \PtDimuon -dependent efficiency (estimated from data) and  acceptance (obtained from simulation). A bin-by-bin correction factor is then calculated as the ratio of the integrals of the rescaled to the original \PtDimuon distributions for each bin.

These correction factors show a mild increase with \Multiplicity. To reduce the statistical fluctuations, a fit is performed with a logistic function to this multiplicity dependence, and the factor used to scale the yields is evaluated at the central \Multiplicity value in every bin. The difference in the ratio between low- and high-multiplicity bins due to the efficiency and acceptance corrections is of the order of 2\%.

The systematic uncertainties due to acceptance and efficiency are calculated by making different choices for their evaluation, and using the new values throughout all the steps of the analysis. For example, alternative procedures are used to estimate the efficiency and acceptance distributions (using simulation instead of collision data for the efficiency calculation, or using different binnings), and the \emph{sPlot} results are compared with those from an invariant mass sideband subtraction method. The only significant effect is found when the mean values of the acceptance and efficiency for all the candidates in a given bin is used instead of the \PtDimuon-linked correction. This gives a systematic variation in the ratio of the order of 1\%.

A final correction to the measured ratios comes from the effect of vertex merging due to pileup. The merging of vertices causes migration of events from lower- to higher-multiplicity bins. It is possible to evaluate the percentage of this migration using simulation. Once a map of the true percentage composition of all the bins is obtained, the ratios can be corrected using an unfolding procedure, starting from the lowest \Multiplicity bin where no merging affects the ratios.
Given that the ratios vary smoothly with \Multiplicity, the final effect is small, and the  largest correction in the highest bin is estimated to be of the order of 1.5\%. Systematic uncertainties from  different pileup conditions and tunings were found to be negligible.

\section{Results and discussion}
\label{sec:results}

\subsection{The Y(nS) ratios vs. multiplicity}
\label{sec:Inclusive}

\begin{figure}[h!]
\centering
   \includegraphics[width=0.45\textwidth]{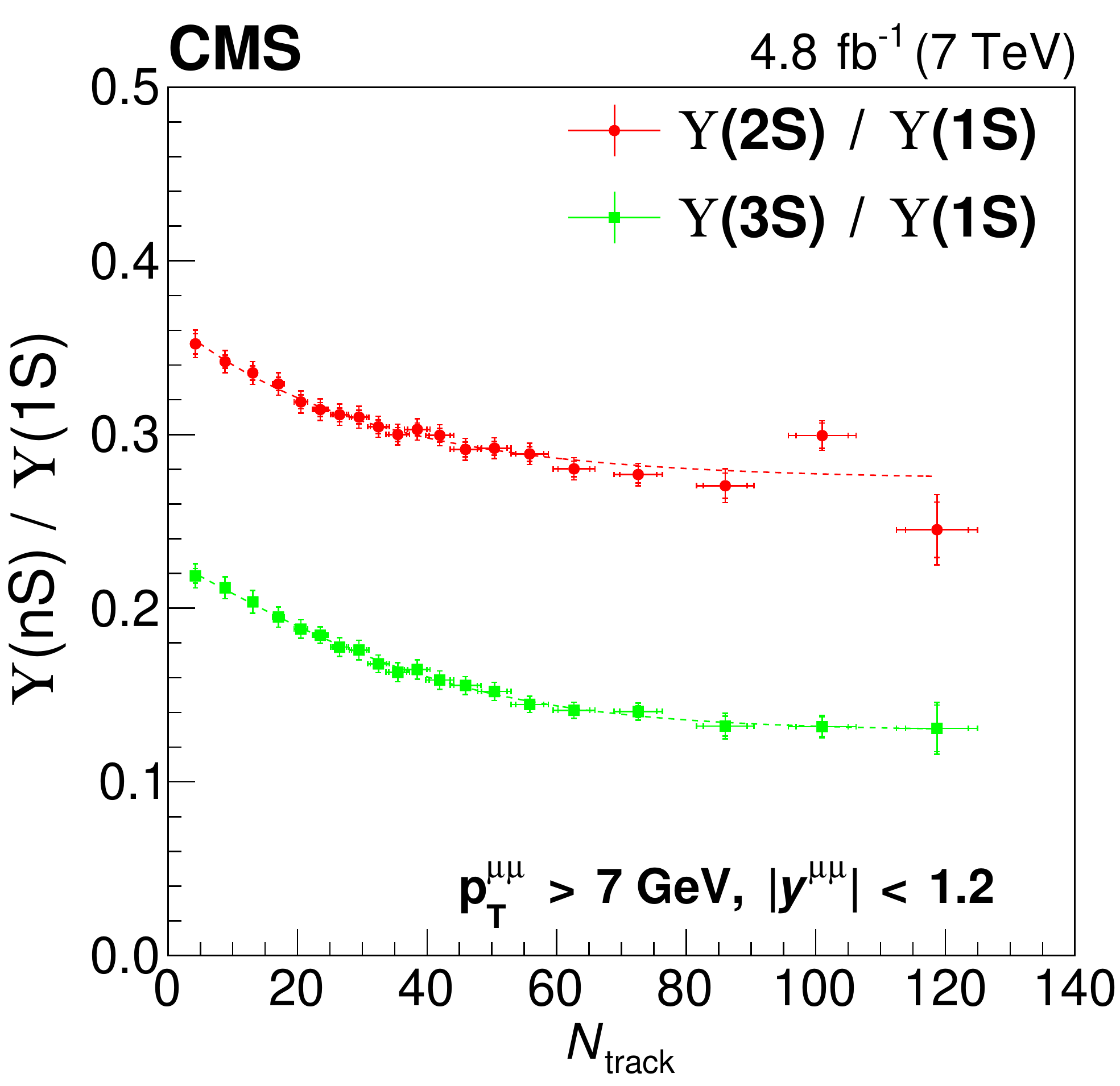}
      \hspace{5mm}
   \includegraphics[width=0.45\textwidth]{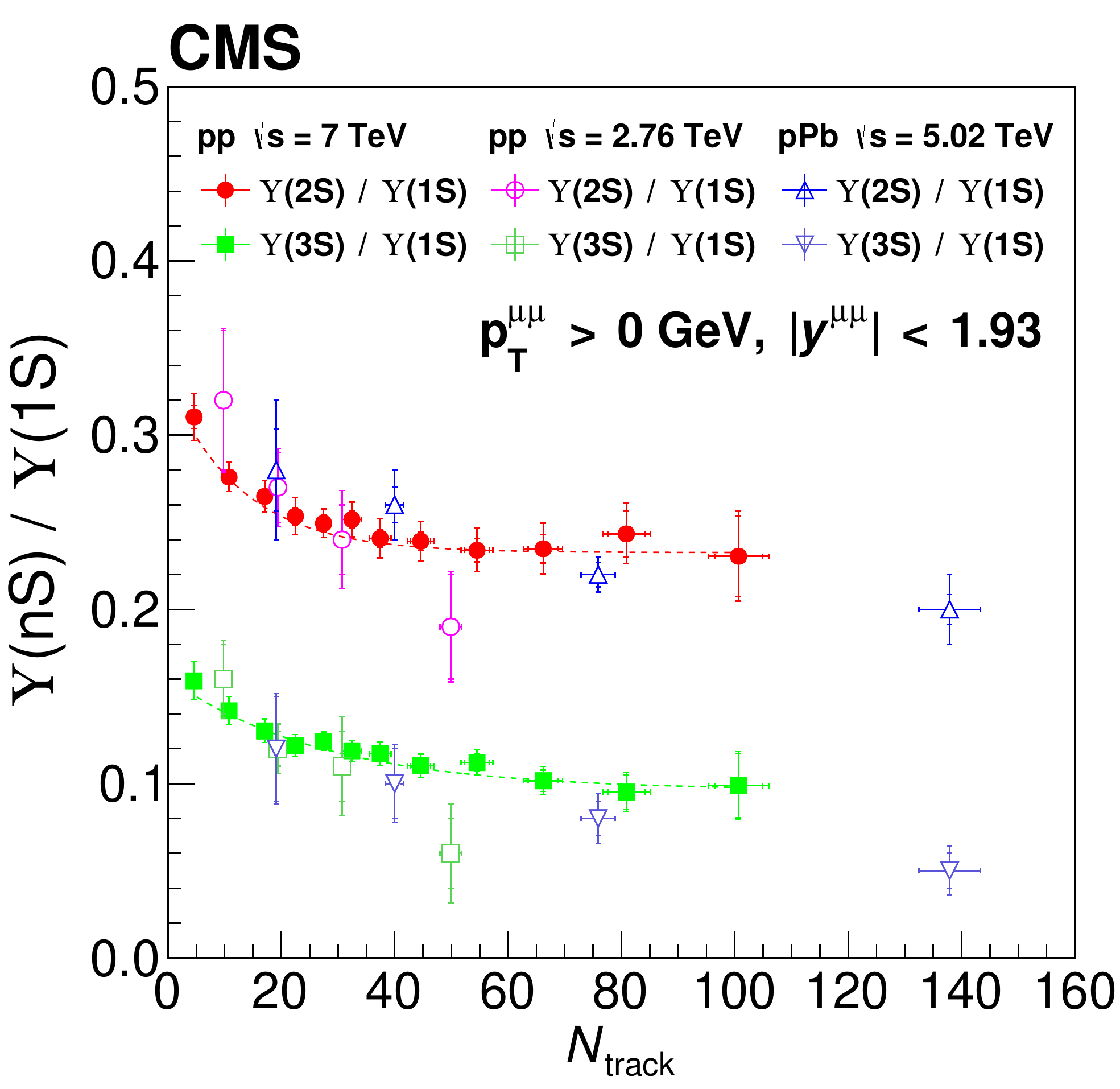}
   \caption{The ratios \RatioA and \RatioB with $\PtDimuon > 7\GeV$ (left) and  $\PtDimuon > 0\GeV$ (right) as a function of \Multiplicity. The lines are fits to the data with an exponential function. The outer vertical bars represent the combined statistical and systematic uncertainties in the ratios, while the horizontal bars give the uncertainty in \MultiplicityMean in each bin. Inner tick marks show only the statistical uncertainty, both in the ratio and in \MultiplicityMean. The results of Ref.~\cite{Chatrchyan:2013nza}  are shown in the right plot for comparison, and  a small correction is applied to the present results to account for the different rapidity ranges in the measurements, $\RapDimuon < 1.20$ here and $\RapDimuon < 1.93$ in Ref.~\cite{Chatrchyan:2013nza}.}
\label{fig:2011sys}
\end{figure}

The measured \RatioA and \RatioB values are shown in Fig.~\ref{fig:2011sys}, as a function of \Multiplicity, for both the (left) $\PtDimuon > 7\GeV$ (4.8\fbinv) and (right) $\PtDimuon > 0\GeV$ (0.3--4.8\fbinv) samples. In Fig.~\ref{fig:2011sys} (right), the CMS results of Ref.~\cite{Chatrchyan:2013nza}  for a smaller $\Pp\Pp$ sample at $\sqrt{s} = 2.76\TeV$ and in $\Pp \mathrm{Pb}$ collisions at 5.02\TeV are overlaid on the current results  for comparison.  In those samples, no \PT cut was imposed on the \PgUn, hence  the smaller sample from this analysis starting at $\PT = 0$ is included. A small 2\% correction is applied to the present results to account for the different rapidity ranges in the three measurements, based on the measured rapidity dependence of the \PgUn production cross sections~\cite{Chatrchyan:2013yna}.

A clear trend is visible in both plots with a decrease in the ratios from low- to high-multiplicity bins. 
The trend is similar in the two kinematic regions, and reminiscent of the measurements from Ref.~\cite{Chatrchyan:2013nza}, in particular of the $\Pp \mathrm{Pb}$ results.
To quantify the decrease, a fit is performed using an exponential function: $e^{(p_0 + p_1 x)}+p_2$, with $p_{0}$, $p_{1}$, and $p_{2}$ as free parameters in the fit. To measure the decrease in the ratios from this analysis, the resulting best fit is evaluated at the centre of the lowest and highest \Multiplicity bins. 
In the $\PtDimuon > 7\GeV$  case, this results in a decrease of $(-22 \pm 3 )$\% for \RatioA and  $(-42 \pm 4 )$\% for \RatioB, where the uncertainties combine the statistical (evaluated at the 95\% confidence level) and systematic (using the upper and lower shifts in the ordinates of the data) uncertainties.

\begin{figure}[h!]
\centering
   \includegraphics[width=0.45\textwidth]{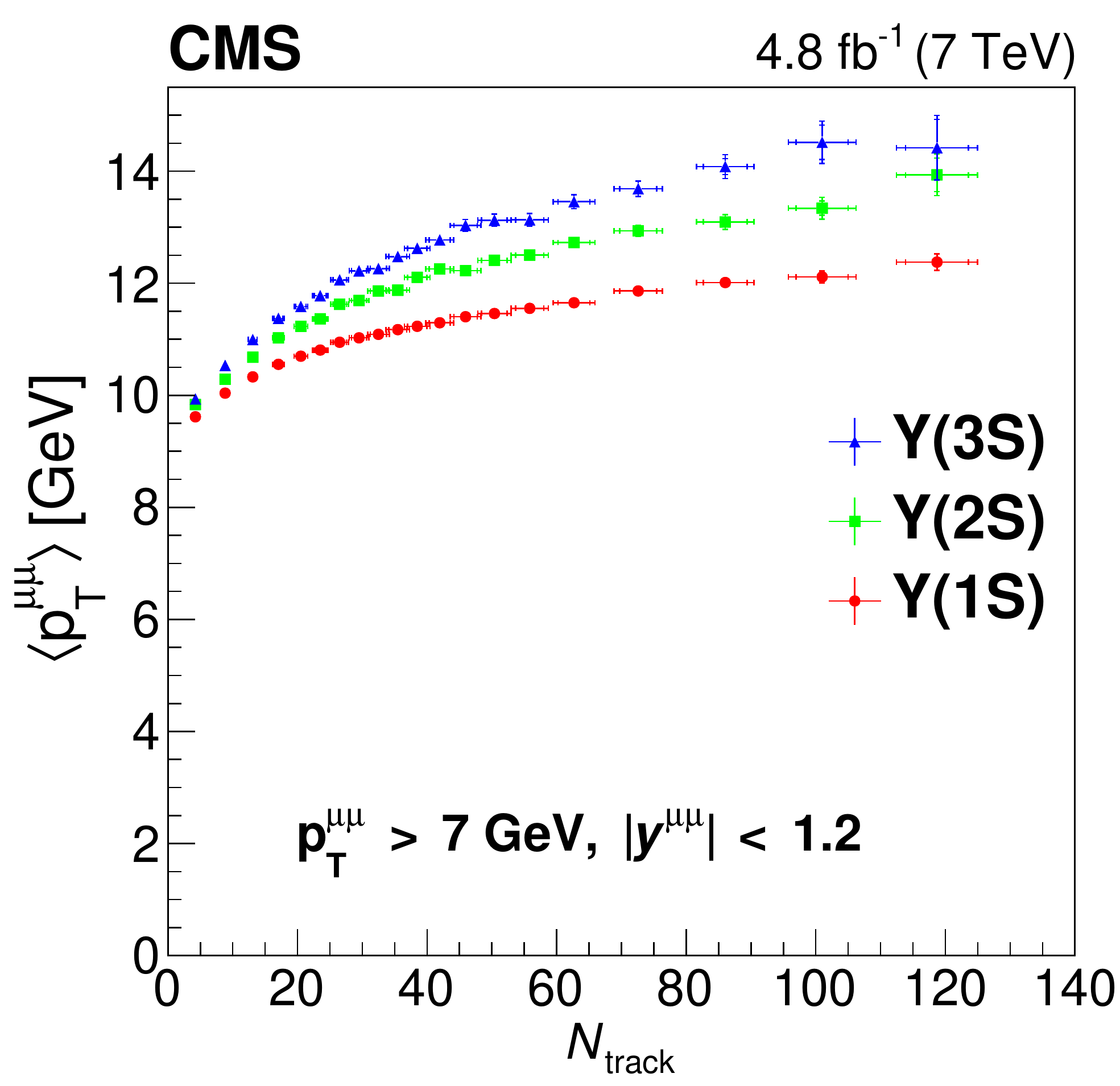}
   \hspace{5mm}
   \includegraphics[width=0.45\textwidth]{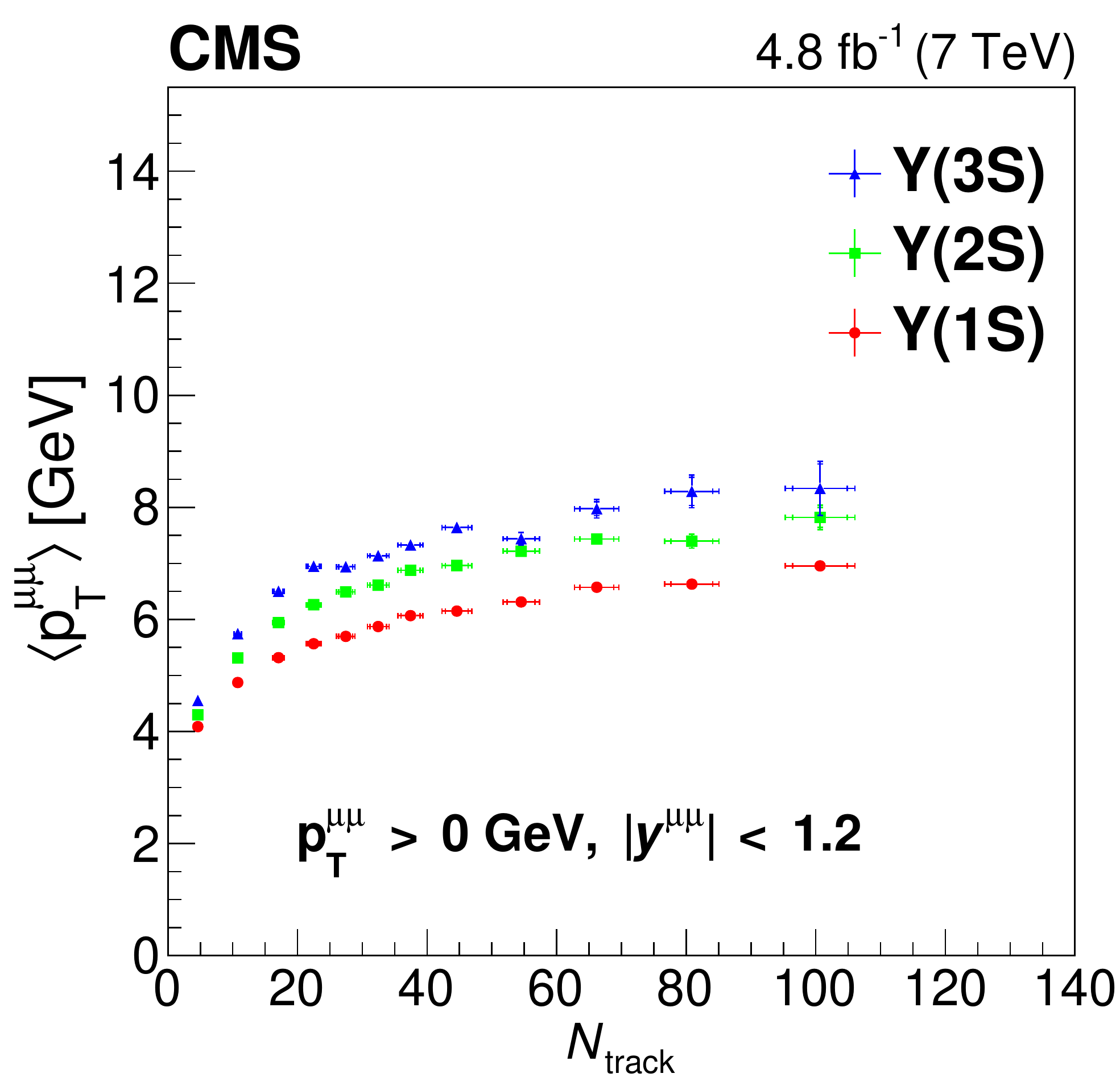}
   \caption{Mean \PtDimuon values for the three \PgUn states as a function of \Multiplicity for $\PtDimuon > 7\GeV$ (left) and $> 0\GeV$ (right). The outer vertical bars represent the combined statistical and systematic uncertainties in the ratios, while the horizontal bars give the uncertainty in \MultiplicityMean in each bin. Inner tick marks show only the statistical uncertainty, both in the ratio and in \MultiplicityMean.}
\label{fig:SpectrumAccCorr2}
\end{figure}

Previous measurements~\cite{Chatrchyan:2013yna} have shown that the ratios \RatioA and \RatioB increase with  \PtDimuon. This effect is also visible in Fig.~\ref{fig:2011sys},
where the values of each ratio are higher in the left plot with a \PtDimuon minimum of 7\GeV than in the right plot with no minimum \PtDimuon requirement. Figure~\ref{fig:SpectrumAccCorr2}~left~(right) shows the mean \PtDimuon values for the three \PgUn states with $\PtDimuon > 7\;(0)\GeV$, as a function of \Multiplicity. This is obtained by taking the \PT spectra of the dimuon candidates using the \emph{sPlot} technique and rescaling them for the efficiency and acceptance corrections as a function of \PtDimuon, as described in Section~\ref{sec:corr}.  From these corrected \PtDimuon distributions the mean value and the corresponding uncertainty are calculated.
We observe a hierarchical structure, where the transverse momentum increases more rapidly with  \Multiplicity as the mass of the corresponding \PgUn increases. An increase with particle mass was also observed in $\Pp\Pp$ collisions at the LHC  for pions, kaons, and protons~\cite{Chatrchyan:2012qb}.

\subsection{Transverse momentum dependence}
\label{sec:TransverseMom}
\begin{figure}[h!]
\centering
  \includegraphics[width=0.45\textwidth]{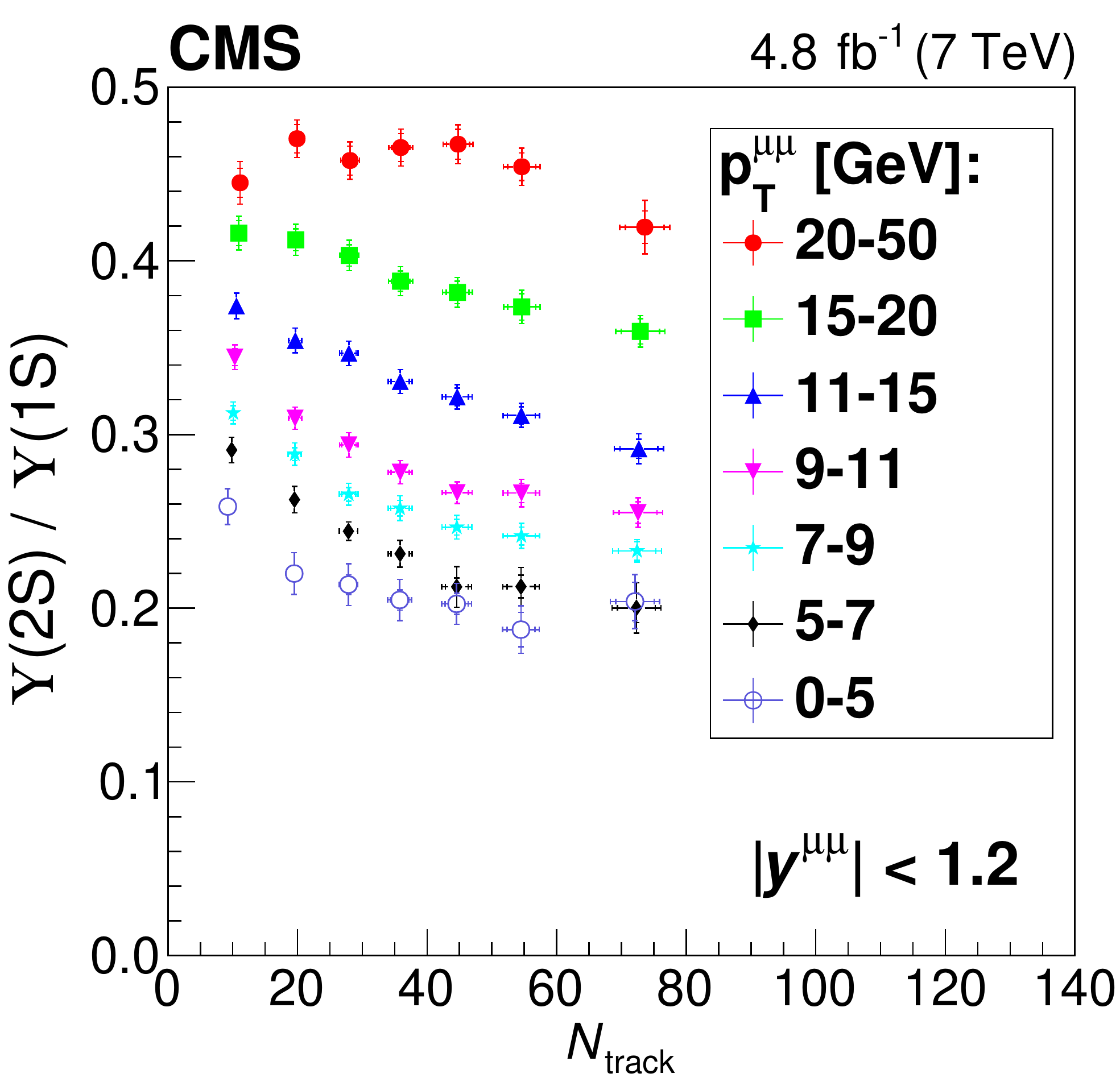}
     \hspace{5mm}
  \includegraphics[width=0.45\textwidth]{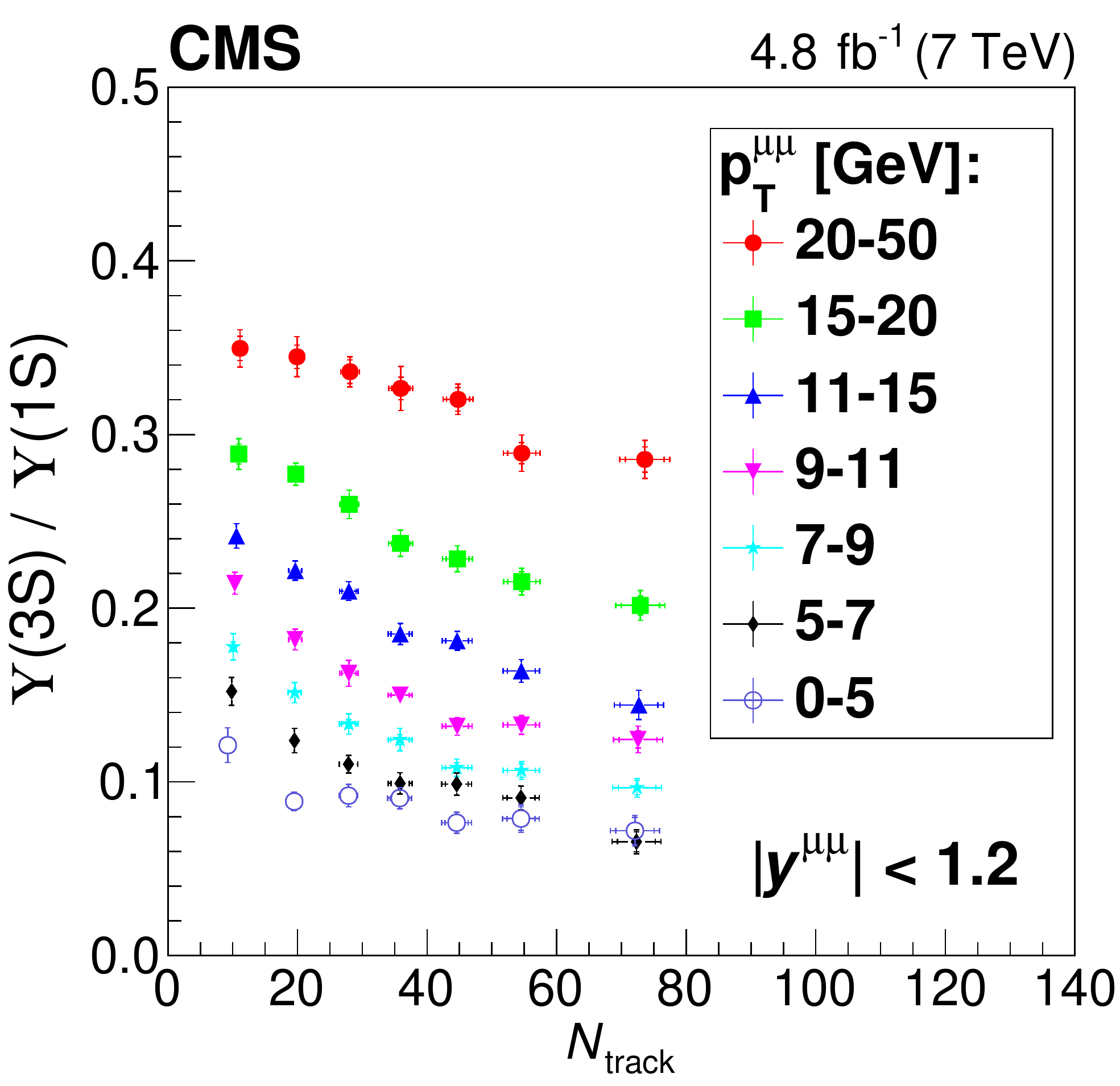}
  \caption{The ratios \RatioA (left) and \RatioB (right) as a function of \Multiplicity,  for different \PtDimuon intervals. The interval 0--5 \GeV corresponds to an integrated luminosity of  0.3\fbinv, the interval 5--7\GeV to 1.9\fbinv, and the rest to the full integrated luminosity of 4.8\fbinv.  The outer vertical bars represent the combined statistical and systematic uncertainties in the ratios, while the horizontal bars give the uncertainty in \MultiplicityMean in each bin. Inner tick marks show only the statistical uncertainty, both in the ratio and in \MultiplicityMean.}
\label{fig:DoublyPT}
\end{figure}

The ratios \RatioA (left) and \RatioB  (right) are plotted in Fig.~\ref{fig:DoublyPT} as a function of \Multiplicity for seven \PtDimuon intervals from 0 to 50\GeV.

In all the \PtDimuon ranges, there is a decrease in the ratios with increasing multiplicity, with the largest rate of decrease in the $\PtDimuon = 5$--7\GeV bin. At higher \PtDimuon values, the decrease in the ratios is smaller.  This is particularly evident for the $\PtDimuon = 20$--50\GeV bin, especially for \RatioA where the ratio is compatible with being constant. In the 0--5\GeV bin, all the decrease occurs at low multiplicity, with the ratios consistent with being flat beyond the first \Multiplicity bin, especially for the ratio \RatioA.

\subsection{Local multiplicity dependence}
\label{sec:Direction}

\begin{figure}[h!]
\centering
   $\vcenter{\hbox{\includegraphics[width=0.45\textwidth]{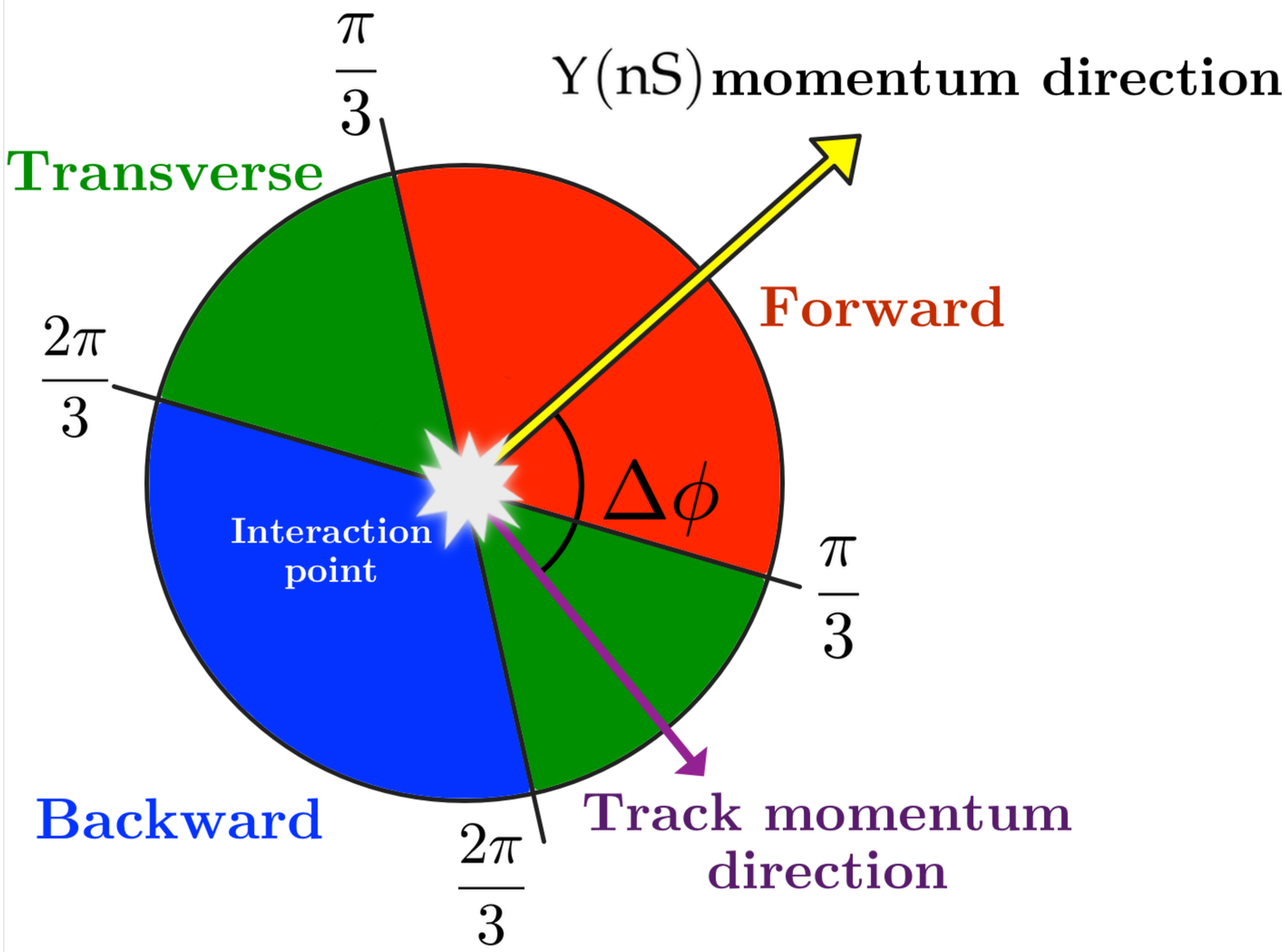}}}$
   \hspace{5mm}
   $\vcenter{\hbox{\includegraphics[width=0.45\textwidth]{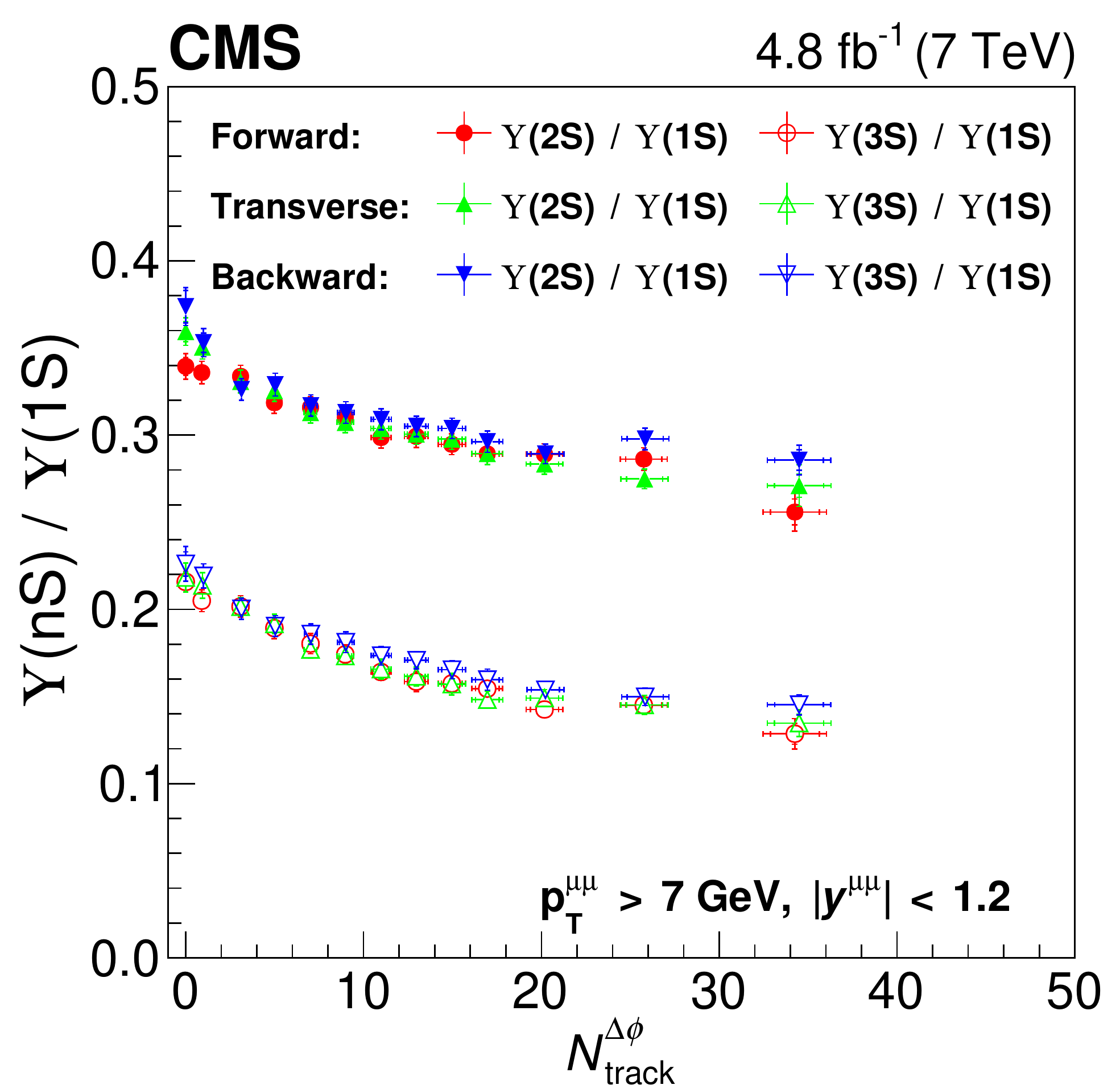}}}$
  \caption{Left: A schematic view in the azimuthal plane of the three $\Delta \phi$ regions with respect to the \PgUn momentum direction. Right: The ratios \RatioA and \RatioB, as a function of \MultiplicityDir for the three $\Delta \phi$ regions shown in the left plot. The outer vertical bars represent the combined statistical and systematic uncertainties in the ratios, while the horizontal bars give the uncertainty in \MultiplicityMeanDir for each bin. Inner tick marks show only the statistical uncertainty, both in the ratio and in \MultiplicityMeanDir.}
\label{fig:angles}
\end{figure}

To better investigate the connection between \PgUn production and the UE properties, a new type of multiplicity, \MultiplicityDir, is defined, based on the  difference between the azimuthal angle of each track and the \PgUn meson, $\Delta \phi$. This  relative angular separation is divided into three ranges (as is done in Ref.~\cite{Khachatryan:2010pv}): a \emph{forward} one comprised of $\abs{\Delta \phi} < \pi / 3 $ radians, a \emph{transverse} one with  $\pi / 3 \leq \abs{\Delta \phi} < 2\pi / 3 $ radians, and a \emph{backward} one of $ 2\pi / 3  \leq \abs{\Delta \phi} \leq \pi $ radians, as shown in Fig.~\ref{fig:angles}~(left).

On average, there are about three more tracks in the \emph{forward} ($14.55\pm 0.05$, including the two muons) and \emph{backward} ($14.83\pm 0.05$) regions than in the \emph{transverse} interval ($11.90\pm 0.05$), where the uncertainties are statistical only.  Similar values are obtained when considering the  \PgUa, \PgUb, and \PgUc~mesons separately.

The \PgUn ratios are presented as a function of \MultiplicityDir  in the three  azimuthal intervals in Fig.~\ref{fig:angles} (right), where the decrease in the ratios is again visible, with similar trends in the three angular regions. The main differences are present at low \MultiplicityDir, where the ratios are slightly higher when considering the backward azimuthal region.
In particular, the fact that the decrease is present in the transverse region suggests its connection with the UE itself, rather than a dependence on the particle activity along the \PgUn direction, which would produce additional particles only in the forward region. The same consideration applies to unaccounted effects coming from feed-down, \ie from \PgUn states not produced in the hard scatter, as discussed in the following section.

\subsection{Dependence on the Y(nS) isolation}
\label{sec:Comover}
The isolation of the \PgUn is defined by the number of tracks found in a small angular region around its direction. The study is aimed at verifying whether charged tracks produced along the \PgU~momentum direction, such as the "comovers" of Ref.~\cite{Ferreiro:2012rq}, could explain the observed reduction in the cross section ratio. The number of particles (\MultiplicityCone) in a cone around the \PgU~momentum direction~($\Delta\mathrm{R} = \sqrt{\smash[b]{(\Delta\eta)^2+(\Delta\phi)^2}} <0.5$) is counted, where $\Delta\eta$ is the difference in pseudorapidity between the \PgUn and the other particles. The data sample is split into four categories: $\MultiplicityCone  = 0, 1, 2,$ and $> 2$. In the last case, for the lower multiplicity range 0--15, a strong decrease in both ratios was initially observed. The source was identified as an enhancement of the  \PgUa signal coming from the feed-down process $\PgUb \to \PgUa \PGpp \PGpm$. This was verified by reconstructing the \PgUb state using the selection and  procedure of Ref.~\cite{Chatrchyan:2013mea}. While the raw number of reconstructed \PgUb  events from the fit to the  $\PgUa \PGpp \PGpm$  mass spectrum is  below 1\% in all the \Multiplicity bins, this component increases significantly, up to 25\%,  when we require tracks in the $\Delta\mathrm{R} <0.5$ cone. On the other hand, the contributions from $\PgUc \to \PgUa \PGpp \PGpm$ and $\PgUc \to \PgUb \PGpp \PGpm$ decays  remain negligible. A correction is applied to take into account both the number of reconstructed feed-down events and the probability that an event is selected in that multiplicity bin due to the presence of the  feed-down $\PGpp \PGpm$ pair. A sizeable  (of the order of 30\%) correction is needed only for the \Multiplicity = 0--15 bin, when requiring more than two particles in the cone. 
The  ratios \RatioA and  \RatioB vs. track multiplicity in the four different categories, after this correction, are shown in Fig.~\ref{fig:ComoverSphericity}~(left). The dependence on the charged particle multiplicity is similar in all the categories and also shows a flattening in the $\MultiplicityCone > 2$  category, which is opposite to what would be expected in the comover picture. 

\subsection{Transverse sphericity dependence}
\label{sec:Sphericity}

The transverse sphericity is a momentum-space variable, useful in distinguishing the dominant physics process in the interaction. It is defined as:
\begin{linenomath}
\begin{equation*}
S_\mathrm{T} \equiv \frac{2{\lambda}_2}{{\lambda}_1+{\lambda}_2},
\end{equation*}
\end{linenomath}
where $\lambda_1 > \lambda_2$ are the eigenvalues of the matrix constructed from the transverse momenta components of the charged particles (labelled with the index \emph{i}), linearised by the additional term $1/p_{\mathrm{T}i}$ (following Ref.~\cite{Abelev:2012sk}):
\begin{linenomath}
\begin{equation*}
S_{xy}^{T}  =  \frac{1}{\sum_i p_{\mathrm{T}i}} \sum_{i}{  \frac{1}{p_{\mathrm{T}i}} 
\begin{pmatrix}
  p_{xi}^2 &  p_{xi} p_{yi} \\
  p_{xi} p_{yi} & p_{yi}^2 \\
 \end{pmatrix}
 }.
\end{equation*}
\end{linenomath}

\begin{figure}[h!]
\centering
   \includegraphics[width=0.45\textwidth]{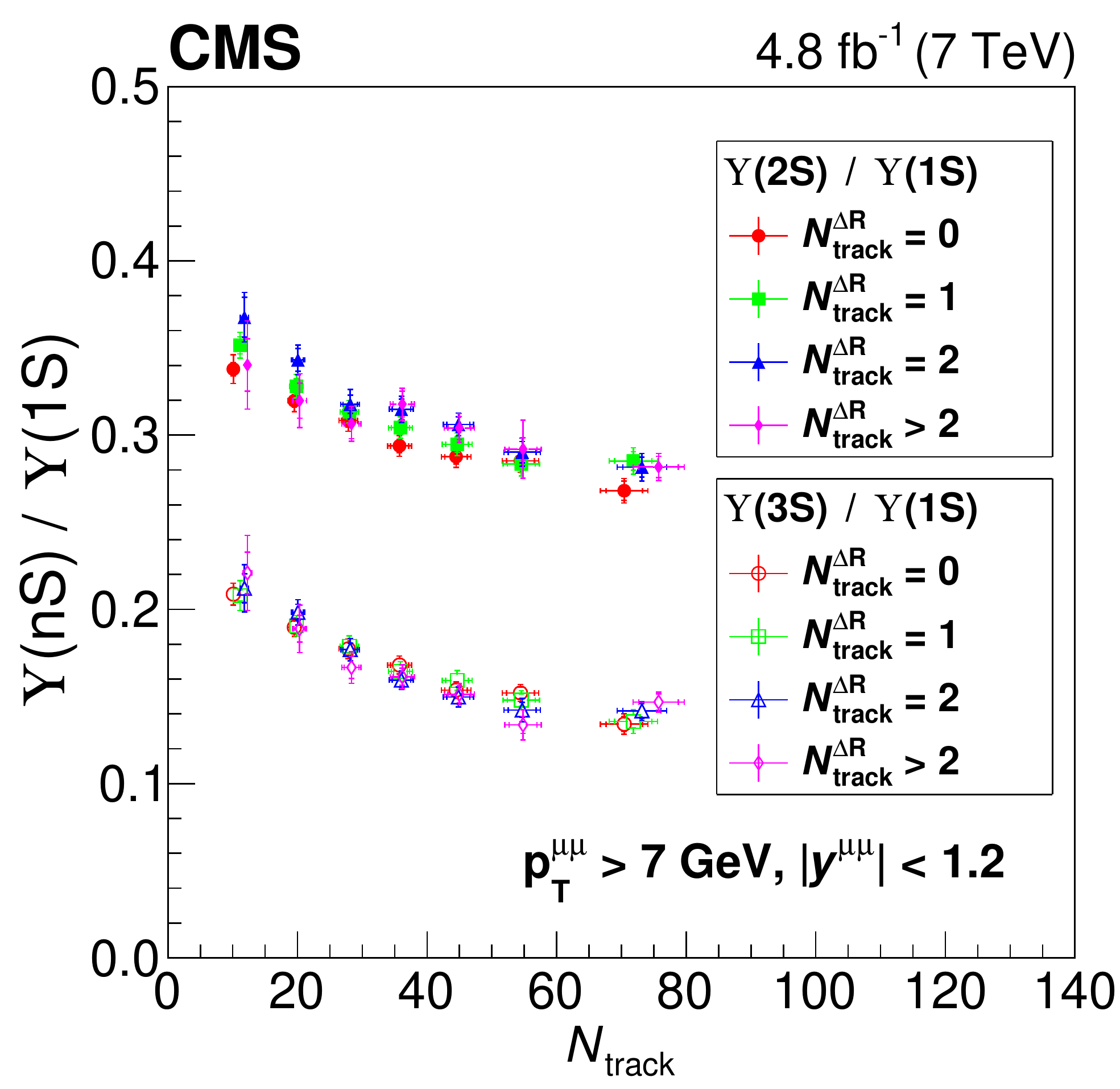}
   \hspace{5mm}
   \includegraphics[width=0.45\textwidth]{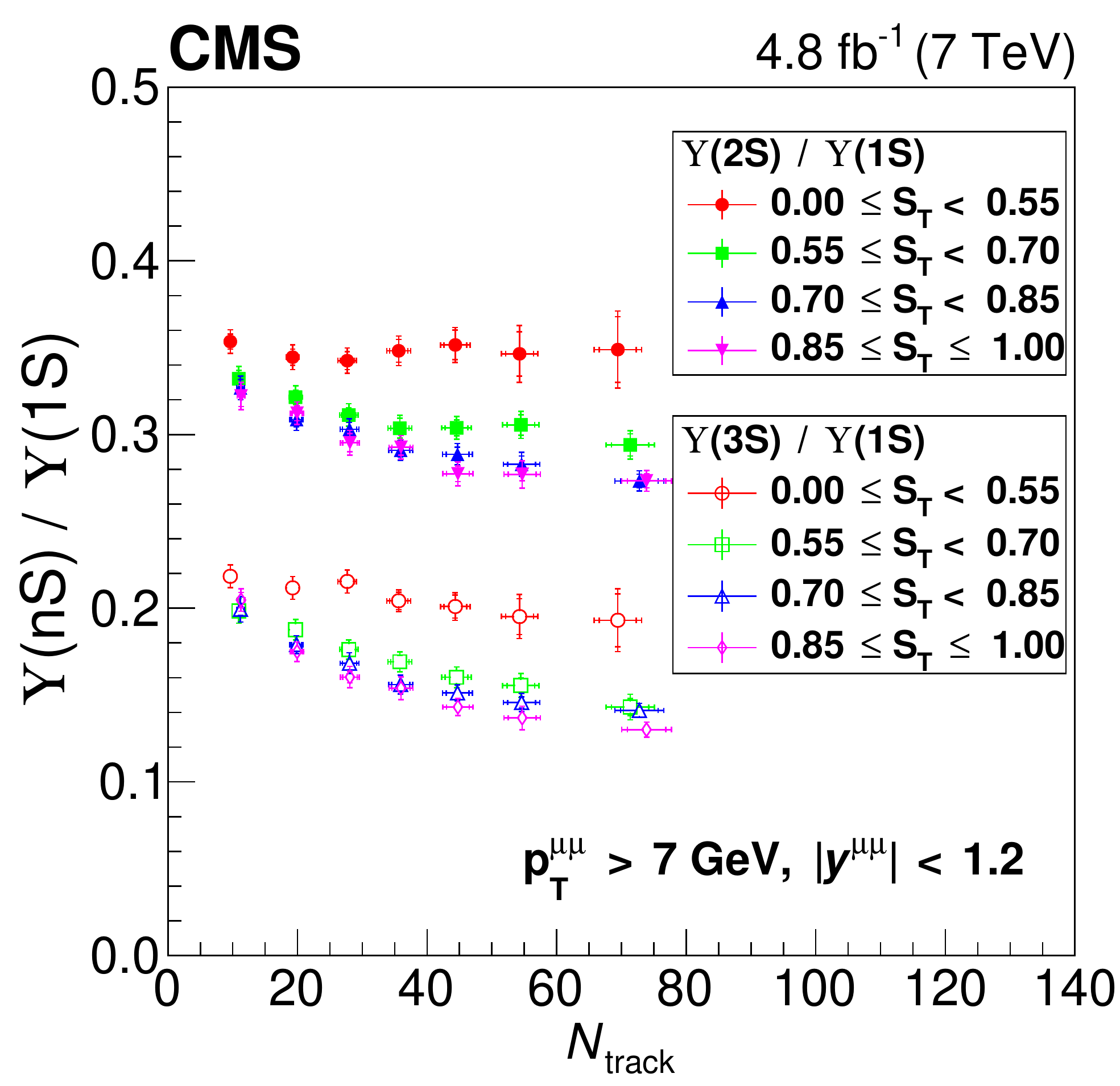}
  \caption{The ratios \RatioA and \RatioB are shown as a function of the track multiplicity \Multiplicity: in four categories based on the number of charged particles produced in a  $\Delta\mathrm{R} < 0.5$ cone around the \PgU~direction  (left), and in different intervals of charged particle transverse sphericity, \ST (right).  The outer vertical bars represent the combined statistical and systematic uncertainties in the ratios, while the horizontal bars give the uncertainty in \MultiplicityMean in each bin. Inner tick marks show only the statistical uncertainty, both in the ratio and in  \MultiplicityMean.}
\label{fig:ComoverSphericity}
\end{figure}

By construction, an isotropic event has sphericity close to 1 ("high" sphericity), while  "jet-like" events have \ST close to zero.  For very low multiplicity,  \ST tends to take low values, so  its definition is inherently multiplicity dependent. 
The cross section ratio between the \PgUn states is evaluated as a function of multiplicity  in four transverse sphericity intervals, 0--0.55, 0.55--0.70, 0.70--0.85, and 0.85--1.00. The resulting trends are shown in Fig.~\ref{fig:ComoverSphericity} (right).
In the low-sphericity region, the ratios remain nearly independent of multiplicity, while the three bins with  $\ST > 0.55$ show a similar decrease as a function of  multiplicity. This observation suggests that the decrease in the ratios is an UE effect. When the high multiplicity is due to the presence of jets or other localised objects and \ST is small, the decrease is absent.  It can also help to explain why the multiplicity dependence is almost flat at higher \PtDimuon, as shown in Fig.~\ref{fig:DoublyPT}. This is because low-sphericity events have a higher \PtDimuon on average.

\subsection{Discussion}
\label{sec:Discussion}
The impact of additional UE particles on the trend of the \PgU cross section ratios to decrease with multiplicity in $\Pp\Pp$ and $\Pp \mathrm{Pb}$ collisions was pointed out in Ref.~\cite{Chatrchyan:2013nza}. In particular, it was noted that the events containing the ground state had about two more tracks on average than the ones containing the excited states. It was concluded that the feed-down contributions cannot solely account for this feature. This is also seen in the present analysis, where the \PgUa meson is accompanied by about one more track on average ($\MultiplicityMean = 33.9 \pm 0.1$) than the \PgUb ($\MultiplicityMean = 33.0 \pm 0.1$), and about two more than the \PgUc ($ \MultiplicityMean =  32.0 \pm 0.1 $). However, as seen in Fig.~\ref{fig:ComoverSphericity}~(left), no significant change is seen when keeping only events with no tracks within a cone along the \PgUn direction.

One could argue that, given the same energy of a parton collision, the lower mass of the upsilon ground state compared to the excited states would leave more energy available for the production of accompanying particles. On the other hand, it is also true that, if we expect a suppression of the excited states at high multiplicity, it would also appear as a shift in the mean number of particles for that state (because events at higher multiplicities would be missing). Furthermore, if we consider only the events with  $0 < \ST <0.55$, where none or little dependence on multiplicity is present, the mean number of charged particles per event is exactly the same for the three \PgU~states ( $\MultiplicityMean = 22.4 \pm 0.1 $). This suggests that the different number of associated particles is not directly linked to the difference in mass between the three states.  

\section{Summary}
\label{sec:Summary}

The measurement of ratios of the $\PgUn \to \PGmp \PGmm$ yields in proton-proton collisions at $\sqrt{s}= 7\TeV$, corresponding to an integrated luminosity of 4.8\fbinv, collected with the CMS detector at the LHC, are reported as a function of the number of charged particles produced with pseudorapidity $\EtaTrack < 2.4$ and transverse momentum $\PtTrack > 0.4\GeV$.  
A significant reduction of the \PgUb/\PgUa~and \PgUc/\PgUa~production ratios is observed with increasing multiplicity. This result confirms the observation made in proton-proton and proton-lead collisions at lower centre-of-mass energy~\cite{Chatrchyan:2013nza}, with increased precision.
The effect is present in different ranges of \PtDimuon, but decreases with increasing \PtDimuon.
For $\PtDimuon > 7\GeV$, different observables are studied in order to obtain a better description of the phenomenon in connection with the underlying event. No variation in the decrease of the ratios is found by changing the azimuthal angle separation of the charged particles with respect to the \PgU~momentum direction. The same applies when varying the number of tracks in a restricted cone around the Y momentum direction.  However, the ratios are observed to be multiplicity independent for jet-like events.
The presented results give for the first time a comprehensive review of the connection between \PgUn production and the underlying event, stressing the need for an improved theoretical description of quarkonium production in proton-proton collisions. 

\begin{acknowledgments}
   We congratulate our colleagues in the CERN accelerator departments for the excellent performance of the LHC and thank the technical and administrative staffs at CERN and at other CMS institutes for their contributions to the success of the CMS effort. In addition, we gratefully acknowledge the computing centres and personnel of the Worldwide LHC Computing Grid for delivering so effectively the computing infrastructure essential to our analyses. Finally, we acknowledge the enduring support for the construction and operation of the LHC and the CMS detector provided by the following funding agencies: BMBWF and FWF (Austria); FNRS and FWO (Belgium); CNPq, CAPES, FAPERJ, FAPERGS, and FAPESP (Brazil); MES (Bulgaria); CERN; CAS, MoST, and NSFC (China); COLCIENCIAS (Colombia); MSES and CSF (Croatia); RIF (Cyprus); SENESCYT (Ecuador); MoER, ERC IUT, PUT and ERDF (Estonia); Academy of Finland, MEC, and HIP (Finland); CEA and CNRS/IN2P3 (France); BMBF, DFG, and HGF (Germany); GSRT (Greece); NKFIA (Hungary); DAE and DST (India); IPM (Iran); SFI (Ireland); INFN (Italy); MSIP and NRF (Republic of Korea); MES (Latvia); LAS (Lithuania); MOE and UM (Malaysia); BUAP, CINVESTAV, CONACYT, LNS, SEP, and UASLP-FAI (Mexico); MOS (Montenegro); MBIE (New Zealand); PAEC (Pakistan); MSHE and NSC (Poland); FCT (Portugal); JINR (Dubna); MON, RosAtom, RAS, RFBR, and NRC KI (Russia); MESTD (Serbia); SEIDI, CPAN, PCTI, and FEDER (Spain); MOSTR (Sri Lanka); Swiss Funding Agencies (Switzerland); MST (Taipei); ThEPCenter, IPST, STAR, and NSTDA (Thailand); TUBITAK and TAEK (Turkey); NASU (Ukraine); STFC (United Kingdom); DOE and NSF (USA).
   
   \hyphenation{Rachada-pisek} Individuals have received support from the Marie-Curie programme and the European Research Council and Horizon 2020 Grant, contract Nos.\ 675440, 752730, and 765710 (European Union); the Leventis Foundation; the A.P.\ Sloan Foundation; the Alexander von Humboldt Foundation; the Belgian Federal Science Policy Office; the Fonds pour la Formation \`a la Recherche dans l'Industrie et dans l'Agriculture (FRIA-Belgium); the Agentschap voor Innovatie door Wetenschap en Technologie (IWT-Belgium); the F.R.S.-FNRS and FWO (Belgium) under the ``Excellence of Science -- EOS" -- be.h project n.\ 30820817; the Beijing Municipal Science \& Technology Commission, No. Z191100007219010; the Ministry of Education, Youth and Sports (MEYS) of the Czech Republic; the Deutsche Forschungsgemeinschaft (DFG) under Germany's Excellence Strategy -- EXC 2121 ``Quantum Universe" -- 390833306; the Lend\"ulet (``Momentum") Programme and the J\'anos Bolyai Research Scholarship of the Hungarian Academy of Sciences, the New National Excellence Program \'UNKP, the NKFIA research grants 123842, 123959, 124845, 124850, 125105, 128713, 128786, and 129058 (Hungary); the Council of Science and Industrial Research, India; the HOMING PLUS programme of the Foundation for Polish Science, cofinanced from European Union, Regional Development Fund, the Mobility Plus programme of the Ministry of Science and Higher Education, the National Science Center (Poland), contracts Harmonia 2014/14/M/ST2/00428, Opus 2014/13/B/ST2/02543, 2014/15/B/ST2/03998, and 2015/19/B/ST2/02861, Sonata-bis 2012/07/E/ST2/01406; the National Priorities Research Program by Qatar National Research Fund; the Ministry of Science and Higher Education, project no. 02.a03.21.0005 (Russia); the Programa Estatal de Fomento de la Investigaci{\'o}n Cient{\'i}fica y T{\'e}cnica de Excelencia Mar\'{\i}a de Maeztu, grant MDM-2015-0509 and the Programa Severo Ochoa del Principado de Asturias; the Thalis and Aristeia programmes cofinanced by EU-ESF and the Greek NSRF; the Rachadapisek Sompot Fund for Postdoctoral Fellowship, Chulalongkorn University and the Chulalongkorn Academic into Its 2nd Century Project Advancement Project (Thailand); the Kavli Foundation; the Nvidia Corporation; the SuperMicro Corporation; the Welch Foundation, contract C-1845; and the Weston Havens Foundation (USA).
\end{acknowledgments}
\bibliography{auto_generated}

\cleardoublepage \appendix\section{The CMS Collaboration \label{app:collab}}\begin{sloppypar}\hyphenpenalty=5000\widowpenalty=500\clubpenalty=5000\vskip\cmsinstskip
\textbf{Yerevan Physics Institute, Yerevan, Armenia}\\*[0pt]
A.M.~Sirunyan$^{\textrm{\dag}}$, A.~Tumasyan
\vskip\cmsinstskip
\textbf{Institut f\"{u}r Hochenergiephysik, Wien, Austria}\\*[0pt]
W.~Adam, F.~Ambrogi, T.~Bergauer, M.~Dragicevic, J.~Er\"{o}, A.~Escalante~Del~Valle, M.~Flechl, R.~Fr\"{u}hwirth\cmsAuthorMark{1}, M.~Jeitler\cmsAuthorMark{1}, N.~Krammer, I.~Kr\"{a}tschmer, D.~Liko, T.~Madlener, I.~Mikulec, N.~Rad, J.~Schieck\cmsAuthorMark{1}, R.~Sch\"{o}fbeck, M.~Spanring, W.~Waltenberger, C.-E.~Wulz\cmsAuthorMark{1}, M.~Zarucki
% ...
\vskip\cmsinstskip
\dag: Deceased\\
1:  Also at Vienna University of Technology, Vienna, Austria\\
% ...
\end{sloppypar}
\end{document}